\documentclass[11pt]{article}
\usepackage{geometry}
\usepackage{dsfont}
\usepackage{epsfig}
\newcommand{\I}{\mathrm{i}}

\begin{document}
\sffamily

\begin{center}
{\LARGE
New canonical and grand canonical DoS techniques \\ 
\vskip2mm
for finite density lattice QCD}

\vskip12mm
Christof Gattringer, Michael Mandl, Pascal T\"orek
\vskip5mm
University of Graz, Institute for Physics\footnote{Member of NAWI Graz.},  A-8010 Graz, Austria 
\vskip12mm

\end{center}

\begin{abstract}
We discuss two new DoS approaches for finite density lattice QCD. The paper 
extends a recent presentation of the new techniques based on Wilson fermions, while here we now discuss and test 
the case of finite density QCD with staggered fermions. The first of our two approaches is based on the canonical 
formulation where observables at a fixed net quark number $N$ are obtained as Fourier 
moments of the vacuum expectation values at imaginary chemical potential $\theta$. We treat the latter as 
densities which can be computed with the recently developed FFA method. The second approach is based on a 
direct grand canonical evaluation after rewriting the QCD partition sum in terms of a suitable pseudo-fermion
representation. In this form the imaginary part of the pseudo-fermion action can be identified and the corresponding
density may again be computed with FFA. We develop the details of the two approaches and discuss some exploratory 
first tests for the case of free fermions where reference results for assessing the new techniques may be obtained from 
Fourier transformation. 
\end{abstract}

\vskip8mm

\section{Introduction}
One of the major open challenges for numerical lattice field theory is the treatment of QCD at finite density. The
central problem is the fact that at finite density the fermion determinant is complex and cannot be used as a 
probability in Monte Carlo simulations. Density of states (DoS) techniques have been among the possible strategies 
for overcoming the complex action problem since the pioneering days of lattice QCD \cite{Gocksch}--\cite{Ejiri}.  
The key challenge for DoS techniques is accuracy, since for computing observables the density needs to be integrated 
over with a highly oscillating factor. A simple sampling of the density with histogram techniques will allow one
to access only very low densities. 

An important step for the further development of DoS techniques was presented in \cite{Langfeld:2012ah} where,
based on ideas from statistical mechanics \cite{WangLandau}, a suitable parameterization of the density combined with 
restricted vacuum expectation values was used to considerably improve the accuracy for the determination of the 
density of states. In a subsequent series of papers this so-called LLR method was developed further and assessed
for several test cases \cite{Langfeld:2013xbf} -- \cite{Francesconi}. A related DoS technique, the so-called 
functional fit approach (FFA) was proposed in \cite{FFA_1} and successfully tested in 
\cite{FFA_2} -- \cite{Z3_FFA_2}. 

However, all these DoS techniques were formulated for bosonic systems and no approach 
to finite density lattice QCD with modern DoS techniques had been presented. Finally, in \cite{GaMaTo} two 
possible formulations of DoS techniques for lattice field theories with fermions were suggested. One of the 
two formulations is the canonical DoS approach (CanDoS) where the density is computed as a function of the 
imaginary chemical potential $\mu \beta = i \theta$. Canonical partition sum and observables are then obtained 
as Fourier moments of the density and the FFA can be used to obtain sufficient accuracy also for the highly 
oscillating integrals for the higher Fourier modes at large net particle numbers. 

The second DoS approach presented in \cite{GaMaTo} is 
a direct grand canonical DoS formulation (GCDoS) based on rewriting the grand canonical partition sum of lattice QCD 
with a suitable pseudo-fermion representation and identifying the imaginary part of the action in this
representation. Subsequently FFA can be applied to evaluate the density as a function of the imaginary
part and again suitable integrals over the density give rise to vacuum expectation values of observables. 

In \cite{GaMaTo} the two new DoS approaches were presented for the formulation of lattice QCD with Wilson fermions
and first tests were presented for free Wilson fermions at finite density. In this paper we now discuss
the CanDos formulation and the direct GCDoS approach for the formulation of lattice QCD with staggered fermions. 
For the CanDos approach we also present some exploratory tests in the free case which allows one to assess the 
accuracy of the method with exact results and to explore parameters of the new techniques.  

\section{The canonical density of states approach}

In this section we present the basic formulation of the canonical DoS approach (CanDos) 
for finite density lattice QCD. 
We stress, however, that the CanDoS approach can easily be implemented for other fermionic theories, e.g., 
theories with 4-fermi interactions generated with auxiliary Hubbard-Stratonovich fields.

\subsection{Canonical ensemble and density of states} 

We study lattice QCD in $d$ dimensions with two degenerate flavors of quarks. 
The canonical partition sum at a fixed net quark  number $N$ is given by
\begin{equation}
Z_N  \; = \; \int_{-\pi}^\pi \frac{d\theta}{2\pi}\int \! \mathcal{D}[U]\,
e^{\,-S_G[U]} \; \det D[U,\mu]^{\,2} \, \bigg|_{\mu\,=\,\I\frac{\theta}{\beta}} \; e^{\, -i \theta N} \; ,
\label{ZN}
\end{equation}
where $S_G[U]$ is the Wilson gauge action (we dropped the constant additive term), 
\begin{equation}
S_G[U] \; = \; -  \, \frac{\beta_G}{3} \sum_{x, \nu < \rho} \mbox{Re} \; \mbox{Tr} \;
U_\nu(x) \, U_\rho(x + \hat{\nu}) \, U_\nu(x+\hat{\rho})^\dagger \, U_\rho(x)^\dagger \; .
\label{Sgauge}
\end{equation}
$\beta_G$ is the inverse gauge coupling and 
the path integral measure $\mathcal{D}[U]$ in (\ref{ZN}) is the product of Haar measures for the link variables 
$U_\nu(x) \in$ SU(3). We have already integrated out the fermions and obtained the fermion determinants for 
the two flavors. $D[U,\mu]$ is the Dirac operator at finite chemical potential 
$\mu$. In this study of the canonical DoS approach we use the staggered Dirac operator, but stress that it is 
straightforward to implement the formalism also for different discretizations of the Dirac operator, e.g., for 
Wilson fermions (compare \cite{GaMaTo}). The staggered Dirac operator $D[U,\mu]$ is given by
\begin{equation}
D[U,\mu]_{x,y} \; = \; m \, \delta_{x,y} \, \mathds{1}_3  \; + \; \frac{1}{2} \sum_{\nu=1}^d \eta_\nu(x) \left[
e^{\, \mu \, \delta_{\nu,d} }  U_\nu(x)\,\delta_{x+\hat{\nu},y} \, - \,
e^{\, -\mu \, \delta_{\nu,d} } U_\nu(x-\hat{\nu})^\dagger\,\delta_{x-\hat{\nu},y} \right] \; ,
\label{dirac_op}
\end{equation}
where $\eta_\nu(x) = (-1)^{x_1 + \, ... \, + x_{\nu-1}}$ are the staggered sign factors and $\mathds{1}_3$ 
is the unit matrix in color space. We work on a $d$-dimensional lattice of size $N_S^{d-1} \times N_T$, 
where the temporal ($\nu = d$) extent $N_T$ gives the inverse temperature in lattices units, i.e., $\beta = N_T$. 
All boundary conditions are periodic, except for the anti-periodic temporal ($\nu = d$) boundary conditions for the 
fermions. $m$ denotes the bare quark mass and $\mu$ the chemical potential. 

In order to project the partition function $Z_N$ to fixed net quark number $N$, 
in (\ref{ZN}) the chemical potential $\mu$ is set to 
$\mu = i \theta/\beta = i \theta / N_T$, and subsequently 
integrated over the angle $\theta$ with a Fourier factor 
$e^{\,-i\theta N}$. This Fourier transformation with respect to the imaginary chemical potential 
sets the net quark number to $N$ and thus generates $Z_N$. The corresponding free energy density 
is defined as $f_N = - \ln Z_N / V$, where $V = N_S^{d-1} N_T$ denotes the $d$-dimensional volume.

Bulk observables and their moments can be obtained as derivatives of $f_N$ with respect to couplings of the theory. 
A simple example, which we also will consider in our numerical tests below, is the chiral condensate 
$\langle \, \overline{\psi}(x) \psi(x) \, \rangle_N = \partial f_N / \partial m$, 
\begin{equation}
\langle \, \overline{\psi}(x) \psi(x) \, \rangle_N \; = \; - \frac{2}{V} 
\frac{1}{Z_N} \int\limits_{-\pi}^\pi \frac{d\theta}{2\pi} \int \! \mathcal{D}[U] \,
e^{\,-S_\mathrm{G}[U]}\; \det D[U,\mu]^{\, 2} \; \mbox{Tr} \, D^{-1}[U,\mu] \, \bigg|_{\mu\,=\,i \frac{\theta}{\beta}} \,
e^{\,- i \theta N} \; .
\label{scalar}
\end{equation}
The mass derivative leads to the insertion of Tr$\, D^{-1}[U,\mu]$ in the path integral. Similarly, 
general vacuum expectation values of some observable ${\cal O}$ at fixed net quark number $N$ have the form
\begin{equation}
\langle  {\cal  O } \rangle_N \; = \; \frac{1}{Z_N} \int\limits_{-\pi}^\pi \frac{d\theta}{2\pi} \int \! \mathcal{D}[U] \,
e^{\,-S_G[U]} \; \det D[U,\mu]^2 \,  {\cal O} [U,\mu] \,  \bigg|_{\mu\,=\,\I\frac{\theta}{\beta}} \; e^{\,- i \theta N} \;.
\label{vevs}
\end{equation}

The partition sum (\ref{ZN}) and the expressions for the vacuum expectation values (\ref{vevs}) can be written 
with suitable densities $\rho^{^{(J)}}\!(\theta)$, which we define as 
\begin{equation}
\rho^{^{(J)}}\!(\theta) \; = \; \int \! \mathcal{D}[U] \,
e^{\,-S_G[U]}\,\det D[U,\mu]^{\,2} \, J[U,\mu]\, \bigg|_{\mu\, = \, i \frac{\theta}{\beta}} \; ,
\end{equation}
where $J[U,\mu]$ is set to $J[U,\mu] = \mathds{1}$ for the partition sum and to $J[U,\mu] = {\cal  O }[U,\mu] $ for 
the vacuum expectation values of observables. With the densities $\rho^{^{(J)}}\!(\theta)$ we may express 
$\langle {\cal O} \rangle_N$ and $Z_N$ as
\begin{equation}
\langle  {\cal  O} \rangle_N \; = \; \frac{1}{Z_N} \int\limits_{-\pi}^\pi\frac{d\theta}{2\pi}\,\rho^{^{( {\cal O})}}\!(\theta)\,
e^{\, - i \theta N} 
\quad , \qquad
Z_N \; = \; \int \limits_{-\pi}^\pi\frac{d\theta}{2\pi}\,\rho^{^{(\mathds{1})}}\!(\theta)\,
e^{\, -i \theta N}\;.
\label{obsdens}
\end{equation}
Note that charge conjugation symmetry can be used to show that $\rho^{^{(\mathds{1})}}\!(\theta)$ is an even function 
such that $\rho^{^{(\mathds{1})}}\!(\theta)$ needs to be determined only in the range $\theta \in [0, \pi]$ which cuts the 
numerical cost in half (see, e.g., \cite{GaMaTo}). A general observable  ${\cal  O }[U,\mu]$ can be decomposed into even and odd parts under 
charge conjugation such that also here the corresponding densities $\rho^{^{(J)}}\!(\theta)$ need to be evaluated only for 
$\theta \in [0, \pi]$.

Having defined the densities $\rho^{^{(J)}}\!(\theta)$ 
and expressed observables in the canonical ensemble as integrals over 
the densities we now have to address the problem of finding a suitable representation of the density and how to 
determine the parameters used in the chosen representation. 

\subsection{Parametrization of the density}

We need to determine the densities $\rho^{^{(J)}}\!(\theta)$ for different operator insertions $J$ as discussed 
in the previous section. For notational convenience, in this section
where we now discuss the parameterization of the densities, we denote all densities as $\rho(\theta)$, 
but stress that we need to determine the parameters of the different 
$\rho(\theta)$ independently for every choice of $J$. 

The densities $\rho(\theta)$ are general functions of $\theta$ in the interval $[0,\pi]$ that for a numerical determination 
we need to describe with only a finite number of parameters. For obtaining a suitable parameterization we 
divide the interval $[0,\pi]$ into $M$ subintervals as,
\begin{equation}
[0,\pi] \; = \; \bigcup_{n=0}^{M-1}I_n\quad,\quad\mbox{with}\quad 
I_n=[\theta_n,\theta_{n+1}]\;,
\label{intervals}
\end{equation}
where $\theta_{0}=0$ and $\theta_M=\pi$. Introducing $\Delta_n = \theta_{n+1}-\theta_n$ for the length 
of the intervals $I_n$, we find
$\theta_n \; = \; \sum_{j=0}^{n-1}\Delta_j\;\; \mbox{for}\;\; n\, = \, 0,1,\, ... \;M$.
For the densities $\rho(\theta)$ we now make the Ansatz 
\begin{equation}
\rho(\theta) \; = \; e^{\, -L(\theta)} \; ,
\label{dens}
\end{equation}
where the $L(\theta)$ are continuous functions that are piecewise linear on the intervals $I_n$. We use the normalization 
$L(0) = 0$, which in turn implies $\rho(0) = 1$. 
Introducing a constant $a_n$ and a slope $k_n$ for the linear function in every interval $I_n$ we may write 
$L(\theta)$ in the form 
\begin{equation}
L(\theta) \; = \; a_n \; + \; k_n \, \big[ \theta-\theta_n \big] \; , \; \; \mbox{for} \; \theta \in I_n \, = \, [\theta_n,\theta_{n+1}] \; .
\end{equation}
Since the functions $L(\theta)$ are normalized to $L(0) = 0$ and are required to be continuous 
we can uniquely determine the constants $a_n$ as functions of the slopes $k_n$ and write $L(\theta)$  
in the following closed form  
\begin{equation}
L(\theta) \;= \; d_n \, + \, \theta  \, k_n 
\; , \quad 
\theta \in I_n
\; , \quad 
d_n \, = \, \sum_{j=0}^{n-1} \big[ k_j -k_n \big]\Delta_j  \; \; \mbox{for}\;\; n \, = \, 0, \, ... \; M \; ,
\label{piecewise}
\end{equation}
and express the densities $\rho(\theta)$ as
\begin{equation}
\rho(\theta) \;= \; A_n \, e^{ \, -  \, \theta \, k_n } \; , \quad 
\theta \in I_n \; , \quad  A_n \; = \; e^{\, -  d_n }\, .
\label{rho_interval}
\end{equation}
Obviously the parameterized density $\rho(\theta)$ 
depends only on the $k_n$, i.e., the set of slopes of the linear pieces in the intervals $I_n$. 
We point out that our parametrization allows one to work with intervals $I_n$ of different sizes $\Delta_n$ such that 
in regions where the density $\rho(\theta)$ varies quickly one may choose small $\Delta_n$, 
while in regions of slow variation one may safe computer time by working with larger $\Delta_n$.

\subsection{Evaluation of the parameters of the density}

To compute the slopes $k_n$ that determine the densities we introduce so-called  
restricted expectation values $\langle \, \theta \, \rangle_n(\lambda)$ that are defined as
\begin{equation}
\langle \, \theta \, \rangle_n(\lambda) \; \equiv \; 
\frac{1}{Z_n(\lambda) }
\int\limits_{\theta_n}^{\theta_{n+1}} \!\! \!d\theta \int \! \! \mathcal{D}[U]\;
e^{\, -S_G[U]} \,\theta\,e^{\; \theta \lambda}\, 
\det D[U,\mu]^{\,2} \, J[U,\mu] \, \bigg|_{\mu \, =\, i\frac{\theta}{\beta}}  \, ,
\label{vevrest}
\end{equation}
where again either $J[U,\mu] = \mathds{1}$ or $J[U,\mu] = {\cal O} [U,\mu]$ is chosen, depending on 
whether the slopes of the density for the partition sum $Z_N$ or the vacuum expectation $\langle {\cal O} \rangle_N$ 
are being computed. The corresponding restricted partition sums $Z_n(\lambda)$ we use in (\ref{vevrest}) are given by 
\begin{equation}
Z_n(\lambda) \; \equiv \; 
\int\limits_{\theta_n}^{\theta_{n+1}} \!\!\! d\theta \int \!\! \mathcal{D}[U] \;
e^{\;-S_G[U]}\; e^{\, \theta  \lambda } \; 
\det D[U,\mu]^{\, 2}  \; J[U,\mu] \, \bigg|_{\mu \, = \, i\frac{\theta}{\beta}} \; .
\label{Zrest}
\end{equation}
In the restricted expectation values $\langle \, \theta \, \rangle_n(\lambda)$ and the partition sum 
$Z_n(\lambda)$ the phase angle $\theta$ is integrated only over the interval $I_n$. 
We have also introduced a free real parameter $\lambda$ which couples to $\theta$ 
and enters in exponential form. Varying this parameter allows one to fully explore the $\theta$-dependence of 
the density in the whole interval $I_n$. Since for imaginary chemical potential 
$\mu \, =\, i \theta/\beta$ the fermion determinant is real and after squaring also positive,  
the expectation values $\langle \, \theta \, \rangle_n(\lambda)$ 
can be evaluated without complex action problem in a Monte Carlo simulation as long as the insertions
$J$ are real and positive (for general insertions $J$ needs to be 
decomposed into pieces that obey positivity). 

The important observation now is that for the parameterization (\ref{rho_interval}) we have chosen for the densities,
$\langle \, \theta \, \rangle_n(\lambda)$ and $Z_n(\lambda)$ can be computed also 
in closed form. Writing the partition sum with the density and then inserting the form (\ref{rho_interval}) one obtains 
\begin{equation}
Z_n(\lambda) \, = \, \int_{\theta_n}^{\theta_{n+1}} \!\! d \theta\,\rho(\theta) \, e^{\, \theta \lambda} 
\, = \, e^{- \, d_n} \!\! \int_{\theta_n}^{\theta_{n+1}} \!\! d \theta \, e^{\, - \theta \, k_n }  e^{\, \theta \lambda}
\, = \, e^{- \, d_n} \,\frac{e^{\,\theta_n [ \lambda-k_n ]}}{\lambda-k_n}
\,\Big(e^{\, \Delta_n[\lambda-k_n]}-1\Big) \; .
\end{equation}
From a comparison of (\ref{Zrest}) and (\ref{vevrest}) one finds that the restricted vacuum expectation value  
$\langle \, \theta \, \rangle_n(\lambda)$ can be computed as the derivative 
$\langle \, \theta \, \rangle_n(\lambda) \; = \; d \, \ln Z_n(\lambda) / d \lambda $, such that also 
$\langle \, \theta \, \rangle_n(\lambda)$ can be found in closed form: 
\begin{equation}
\langle \, \theta \, \rangle_n(\lambda) \; \equiv \; \frac{d \ln Z_n(\lambda)}{ d \, \lambda}  \; = \; 
\theta_n+\frac{\Delta_n}{1-e^{\, -\Delta_n[\lambda-k_n]}}
-\frac{1}{\lambda-k_n} \;.
\label{vevtheta}
\end{equation}
Using a multiplicative and an additive normalization we bring 
$\langle \, \theta \, \rangle_n(\lambda)$ into a standard form $V_n(\lambda)$ where the result is
expressed in terms of a simple function $h(s)$,
\begin{equation}
V_n(\lambda) \, \equiv \, \frac{ \langle \, \theta \, \rangle_n(\lambda) -\theta_n}{\Delta_n}
- \frac{1}{2} \, = \, h\big(\Delta_n [\lambda-k_n] \big) \quad \mbox{with} \quad 
h(s) \, \equiv \,  \frac{1}{1-e^{-s}}-\frac{1}{s}-\frac{1}{2} \; .
\label{vtilde}
\end{equation}
The function $h(s)$ obeys $h(0)=0$, $h^\prime(0)=1/12$, and $\lim_{s\to\pm\infty}h(s)=\pm 1/2$. 

The determination of the slope $k_n$ for the interval $I_n$ now consists of the following steps:  
For several values of $\lambda$ one computes the corresponding restricted vacuum expectation values 
$\langle \, \theta \, \rangle_n(\lambda)$ defined in (\ref{Zrest}) and brings them 
into the normalized form $V_n(\lambda)$ defined in Eq.~(\ref{vtilde}). Fitting the corresponding data with 
$h\big(\Delta_n[\lambda-k_n]\big)$ allows one to determine the $k_n$ from a simple stable 
one-parameter fit. From the sets of the slopes $k_n$ we can determine the densities $\rho(\theta)$ 
using (\ref{piecewise}) and (\ref{rho_interval}) and finally compute the observables via the integrals 
(\ref{obsdens}).

\section{An exploratory test of the canDoS approach in the free case}

As a first assessment of the new canonical density of states approach we test the new method for the case of free 
fermions at finite density in two dimensions. This serves to verify the method and the program, and allows for 
exploring the parameters of the method, such as the number of intervals $I_n$ and suitable choices for the
values of $\lambda$. 
In addition, for the free case all steps of the CanDoS approach can be cross-checked with exact results obtained 
from Fourier transformation. 

\subsection{Setting and reference results from Fourier transformation} 

For this first test we use the chiral condensate at fixed particle number 
$\langle \, \overline{\psi}(x) \psi(x) \, \rangle_N = \partial f_N / \partial m$ as our main observable. 
For the free case the corresponding expression (\ref{scalar}) reduces to 
\begin{equation}
\langle \, \overline{\psi}(x) \psi(x) \, \rangle_N \; = \; - \frac{2}{V} 
\frac{1}{Z_N} \int\limits_{-\pi}^\pi \frac{d\theta}{2\pi} \; 
\det D[\mu]^{\, 2} \; \mbox{Tr} \, D^{-1}[\mu] \, \bigg|_{\mu\,=\,i \frac{\theta}{\beta}} \,
e^{\, -i \theta N} \; ,
\label{scalar0}
\end{equation}
where the all links in the Dirac operator (\ref{dirac_op}) were set to $U_\nu(x) = \mathds{1}$. 
For implementing the CanDoS approach for the condensate we need the two densities, 
\begin{equation} 
\rho^{^{(\mathds{1})}}\!(\theta) \; = \;  \det D[\mu]^{\,2} \, 
\bigg|_{\mu\, = \, i \frac{\theta}{\beta}} \quad \mbox{and} \quad 
\rho^{^{(\mbox{\scriptsize Tr}\, D^{-1})}}\!(\theta) \; = \; \det D[\mu]^{\,2} \, \mbox{Tr} \, D^{-1}[\mu] \, 
\bigg|_{\mu\, = \, i \frac{\theta}{\beta}}  \; .
\label{densities_free}
\end{equation}
For determining the slopes $k_n$ of these two densities we thus have to compute the restricted expectation 
values (\ref{vevrest}) for $J = \mathds{1}$ and $J =  \mbox{Tr} \, D^{-1}$. Normalizing the corresponding Monte 
Carlo data according to (\ref{vtilde}) and fitting them with  $h\big(\Delta_n [\lambda-k_n] \big)$ gives rise to the slopes
$k_n$. From the respective sets of slopes we find the densities $\rho^{^{(\mathds{1})}}\!(\theta)$ and 
$\rho^{^{(\mbox{\scriptsize Tr}\, D^{-1})}}\!(\theta)$ using (\ref{piecewise}), (\ref{rho_interval})
and finally the vacuum expectation value $\langle \, \overline{\psi}(x) \psi(x) \, \rangle_N$  
is obtained as
\begin{equation}
\langle \, \overline{\psi}(x) \psi(x) \, \rangle_N \; = \; - \frac{2}{V}
\frac{1}{Z_N} \int\limits_{-\pi}^\pi\frac{d\theta}{2\pi}\,\rho^{^{(\mbox{\scriptsize Tr}\, D^{-1})}}\!(\theta)\,
e^{\, - i \theta N} 
\quad , \qquad
Z_N \; = \; \int \limits_{-\pi}^\pi\frac{d\theta}{2\pi}\,\rho^{^{(\mathds{1})}}\!(\theta)\,
e^{\, -i \theta N}\;.
\label{scalar_rho}
\end{equation}

In the free case reference results can be obtained with the help of Fourier transformation. 
Furthermore, for the case of two flavors in two dimensions, which we are using for our test, we can explore 
the relation $\det D[\mu]^2 = \det D_{naive}[\mu]$ between the determinant of the staggered Dirac operator 
$D[\mu]$ and the determinant of the naive Dirac operator $D_{naive}[\mu]$, which in two dimensions is given by
\begin{equation}
D_{naive}[\mu]_{x,y} = m \, \delta_{x,y} \, \mathds{1}_2 \times \mathds{1}_3  \; + \; 
\frac{1}{2} \sum_{\nu=1}^2 \sigma_\nu \times \mathds{1}_3 \left[
e^{\, \mu \, \delta_{\nu,2} }  \,\delta_{x+\hat{\nu},y} \, - \,
e^{\, -\mu \, \delta_{\nu,2} } \,\delta_{x-\hat{\nu},y} \right] \; ,
\label{dirac_naive}
\end{equation}
where $\sigma_1$ and $\sigma_2$ are the first two Pauli matrices acting on the Dirac indices of 
the two-component spinors used in the naive discretization and $\mathds{1}_2$ is the corresponding 
unit matrix.  All link variables are set to their trivial values, i.e., they are replaced by the $3\times3$ unit matrix
$\mathds{1}_3$. 
The determinant of the naive Dirac operator can be computed by first diagonalizing $D_{naive}[\mu]$ in space-time 
with the help of Fourier transformation and then taking the product of the corresponding 
momentum space Dirac operator determinants over all momenta. 

The density $\rho^{^{(\mathds{1})}}\!(\theta)$ then is simply obtained via numerically evaluating 
$\det D_{naive}[\mu]$ for $\mu = i \theta/\beta$. For the 
density $\rho^{^{(\mbox{\scriptsize Tr}\, D^{-1})}}\!(\theta)$ one may use 
Jakobi's formula ($d \, \det M /dx = \det M \, \mbox{Tr} [ M^{-1} \, dM/dx]$)
for the derivative of a determinant and the fact that $d D/dm = \mathds{1}$ to obtain
\begin{equation}
\rho^{^{(\mbox{\scriptsize Tr}\, D^{-1})}}\!(\theta) \; = \; 
\det D[\mu]^{\, 2} \; \mbox{Tr} \, D^{-1}[\mu] \bigg|_{\mu\, = \, i \frac{\theta}{\beta}} \; = \; 
\frac{1}{2} \, \frac{d}{d m}  \det D[\mu]^{\, 2} \bigg|_{\mu\, = \, i \frac{\theta}{\beta}} \; = \;
\frac{1}{2} \,  \frac{d}{d m} \det D_{naive} [\mu] \bigg|_{\mu\, = \, i \frac{\theta}{\beta}} \; .
\end{equation}
The vacuum expectation value $\langle \, \overline{\psi}(x) \psi(x) \, \rangle_N$ can be obtained
from (\ref{scalar_rho}) by numerically integrating over $\theta$. For the reference data in the plots below
we implemented this integration with Mathematica. 

\subsection{Numerical results for CanDos in the free case}

Having discussed the observables and the corresponding densities for the free case, as well as the evaluation of 
reference data with the help of Fourier transformation, we now come to a brief exploratory numerical test for 
the free case in $d = 2$ dimensions. The results in the plots below were computed on $16 \times 16$ lattices 
at a mass parameter of $m = 0.1$. We used 50 intervals $I_n$ of equal size to parameterize the density in 
the range $[0,\pi]$. For each interval we computed the restricted expectation values
(\ref{vevtheta}) for 20 different values of $\lambda$ using Monte Carlo simulations based on $10^6$ measurements,
where in the simulation 
the fermion determinant was evaluated exactly with Fourier transformation. The restricted expectation
values were then normalized to the form (\ref{vtilde}) and the slopes $k_n$ determined from the corresponding 
fits with $h\big(\Delta_n[\lambda-k_n]\big)$. From the slopes the densities were computed using 
$(\ref{piecewise})$ and $(\ref{rho_interval})$.

\begin{figure}[t!!]
\begin{center}
\includegraphics[height=63mm,clip]{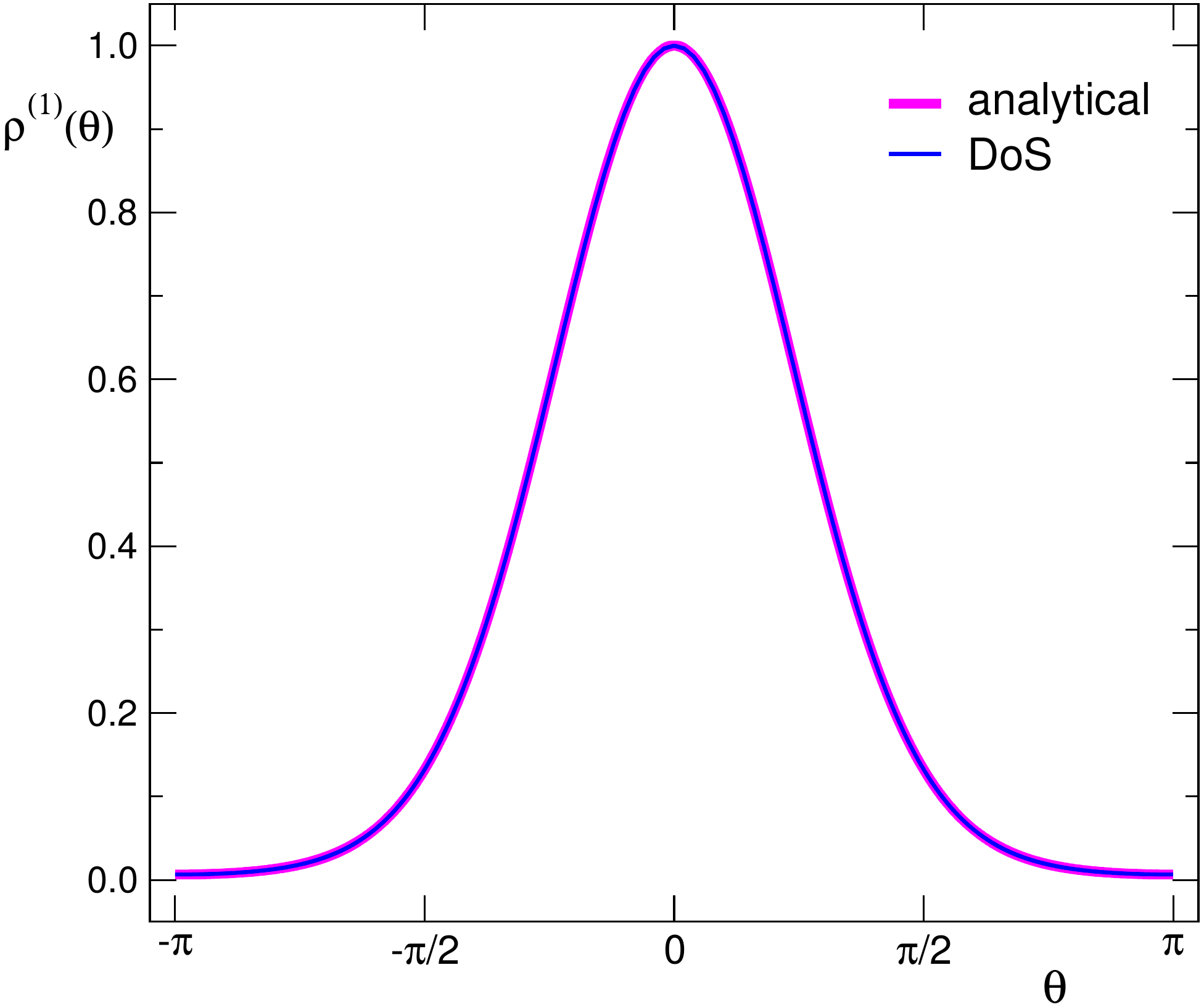}
\includegraphics[height=63mm,clip]{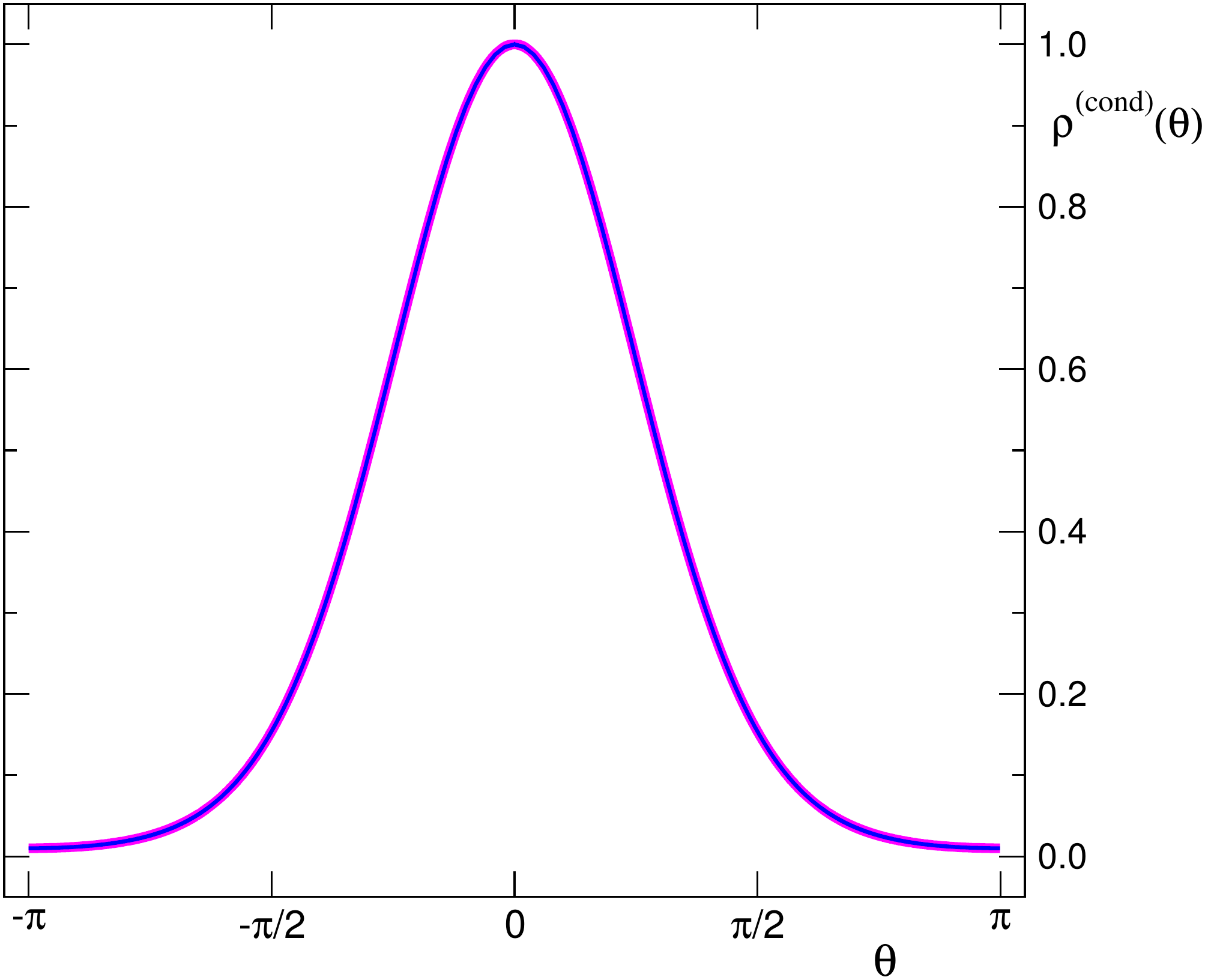}
\end{center}
\vspace{-2mm}
\caption{The densities $\rho^{(\mathds{1})}(\theta)$ (lhs.) and $\rho^{(\,\mbox{\scriptsize Tr} \,D^{-1})}(\theta)$ 
(rhs.\ figure; denoted as $\rho^{(\mbox{\scriptsize cond})}(\theta)$ in the plot). We compare the data from the 
CanDoS determination (thin blue curves) to the analytic results obtained with Fourier transformation (thick magenta
curves). The data are for $16 \times 16$ lattices with $m = 0.1$ and the densities are normalized to 
 $\rho(0)=1$.
\label{fig:denscomp}}
\vspace{8mm}
\begin{center}
\hspace*{-3mm}
\includegraphics[height=63.2mm,clip]{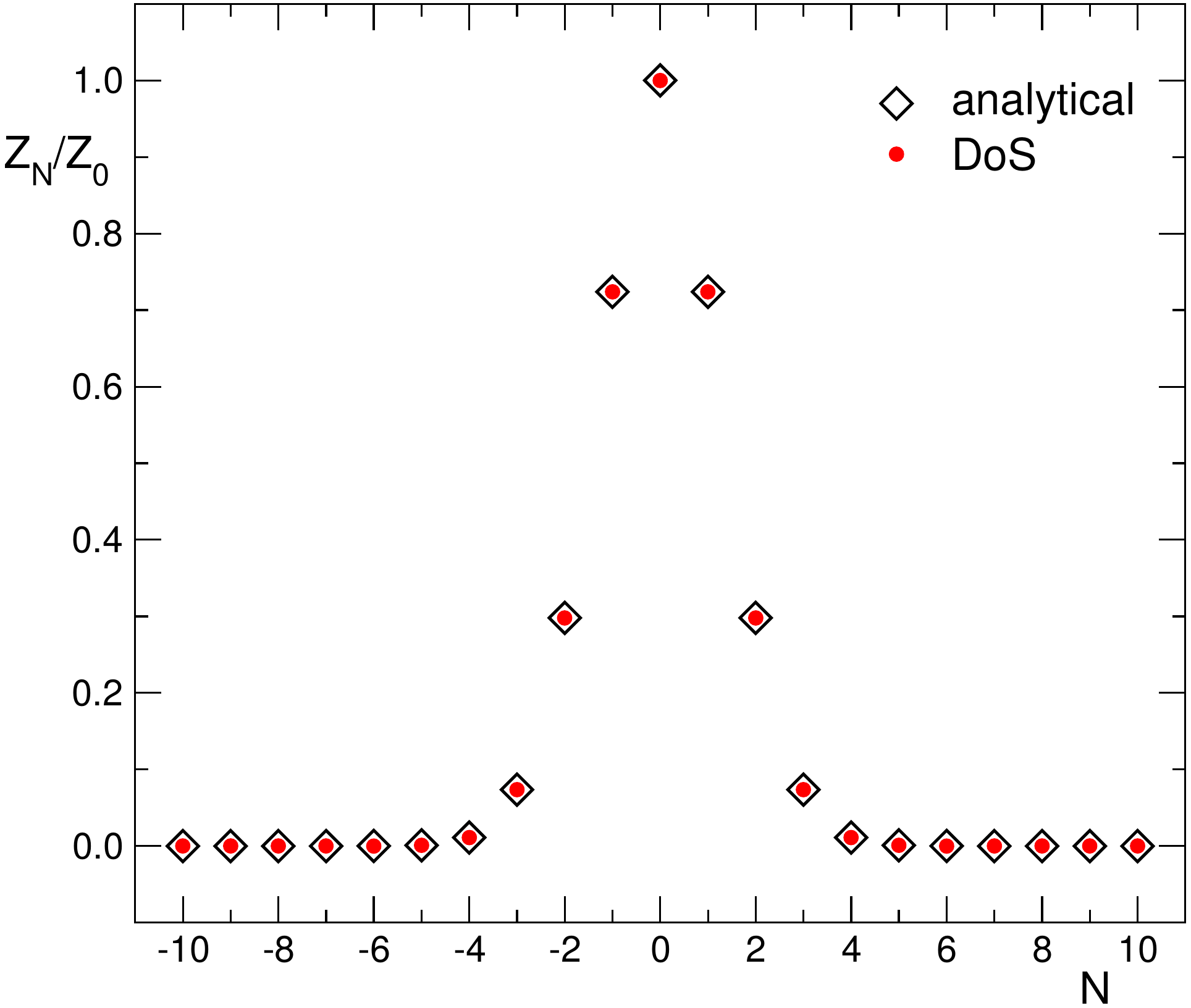}
\includegraphics[height=63.2mm,clip]{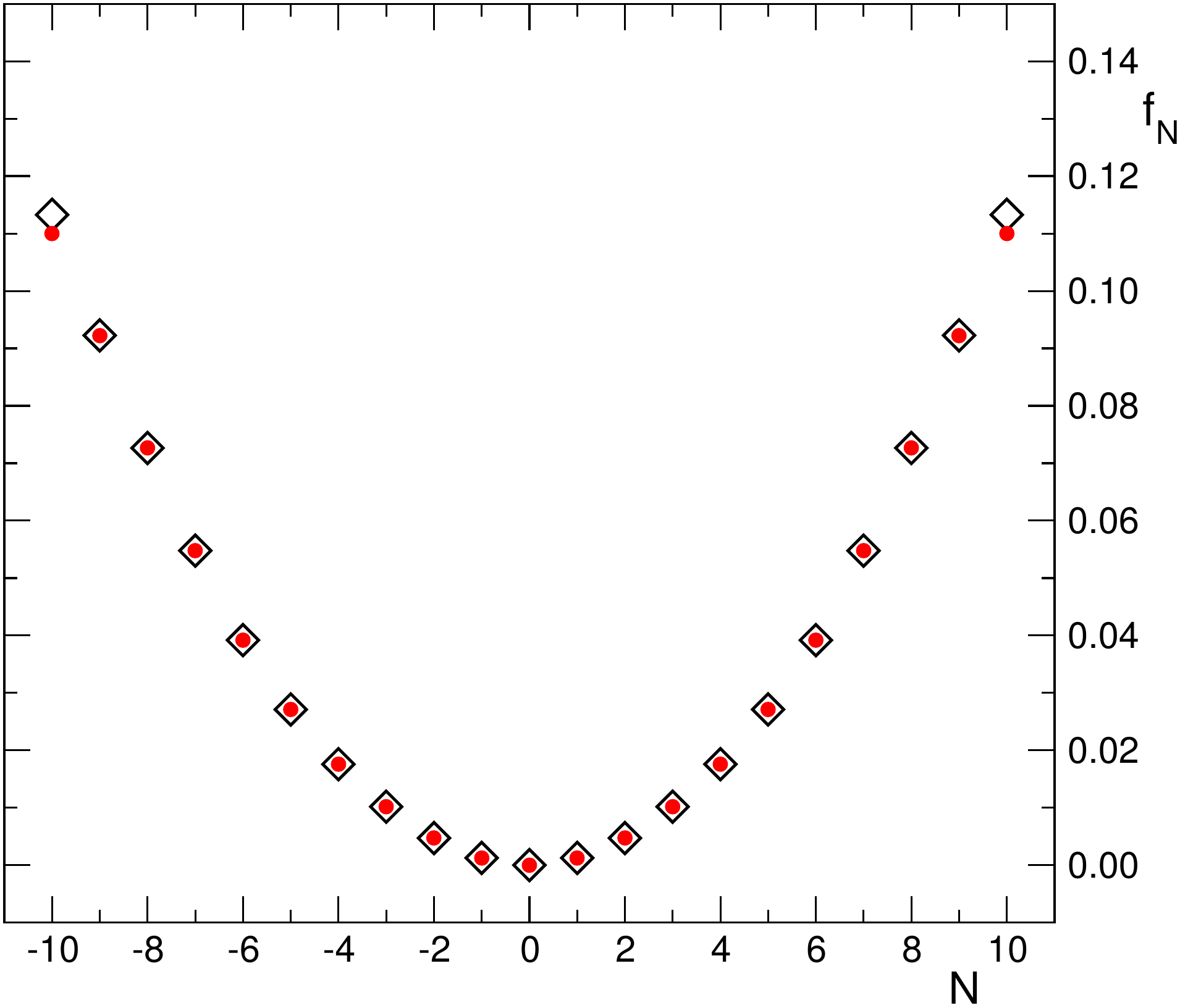}
\end{center}
\vspace{-2mm}
\caption{The canonical partition sums $Z_N/Z_0$ (lhs.) and the corresponding free energy 
densities $f_N = - \ln (Z_N/Z_0) / V$ (rhs.) as a function of the net fermion number $N$.
The parameters are $V = 16 \times 16$ with $m = 0.1$ and we compare the results from the 
CanDoS determination (red dots) to the analytic results obtained with Fourier transformation (black diamonds).
\label{fig:zn}}
\end{figure}

In Fig.~\ref{fig:denscomp} we show the results for the densities $\rho^{(\mathds{1})}(\theta)$ (lhs.\ plot) and 
$\rho^{(\,\mbox{\scriptsize Tr} \,D^{-1})}(\theta)$ (rhs.). The thin blue curves are the results from the 
CanDos determination and the thick magenta curves the reference data computed with Fourier transformation as
discussed in the previous subsection. Obviously the CanDos densities match the reference data very well.

Having determined the densities we can compute the canonical partitions sums $Z_N$ and vacuum expectation values
at fixed net fermion number using (\ref{obsdens}). In the lhs.\ plot of Fig.~\ref{fig:zn} we show our results 
for the canonical partition sums $Z_N$ normalized by $Z_0$ as a function of $N$. The results from the CanDos 
determination are shown as red dots, the reference data from Fourier transformation as black diamonds. Also here 
we observe essentially perfect agreement for all values of the net fermion number $N$ we consider. A more physical 
quantity is the corresponding free energy density $f_N = - \ln Z_N / V$ (here normalized to $f_0 = 0$), which
in the rhs.\ plot of Fig.~\ref{fig:zn} we show as a function of $N$. Again we compare the CanDos results
(red dots) to the corresponding reference data (black diamonds) and find very good agreement and only for the 
largest net particle number $N = 10$ shown in the plot we observe a slight deviation, indicating that for 
net quark numbers $N > 10$ the accuracy of the determination of the density would have to be improved,
e.g., by using more and finer intervals $I_n$.

\begin{figure}[t!!]
\begin{center}
\includegraphics[height=60mm,clip]{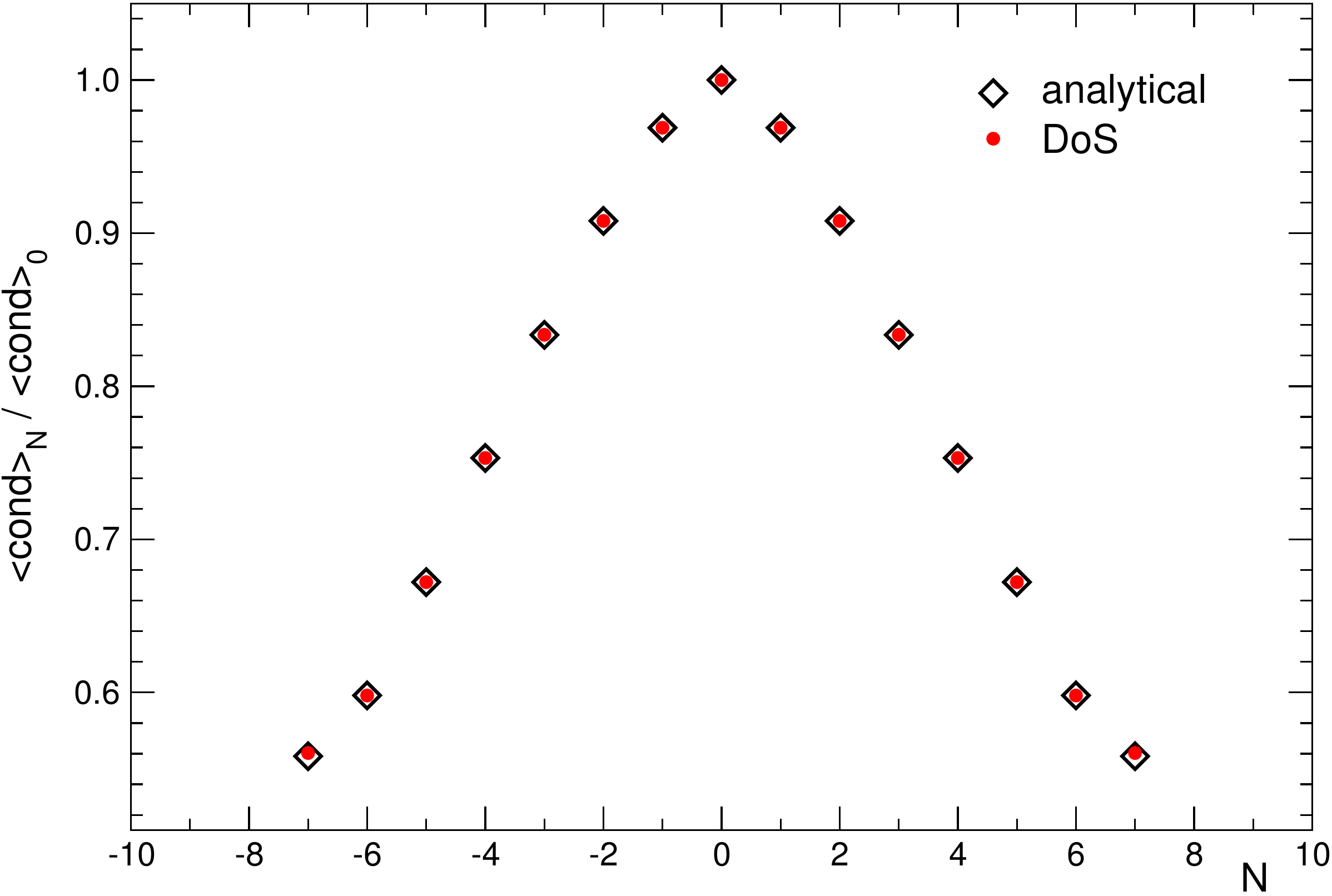}
\end{center}
\caption{The chiral condensate $\langle \overline{\psi}(x) \psi(x) \rangle_N$ 
(in the plot denoted as $\langle cond \rangle_N$ and normalized by the $N = 0$ value) as a function 
of the net fermion number $N$. The parameters are $V = 16 \times 16$ with $m = 0.1$ and we 
compare the results from the 
CanDoS determination (red dots) to the analytic results obtained with Fourier transformation (black diamonds).
\label{fig:condensate}}
\end{figure}

We conclude our exploratory study with discussing the vacuum expectation value of an observable, i.e., 
a case where the ratio of two integrals over two different densities needs to be computed. The quantity we
consider is the chiral condensate and the two corresponding densities 
$\rho^{(\mathds{1})}(\theta)$ and $\rho^{(\,\mbox{\scriptsize Tr} \,D^{-1})}(\theta)$ are the ones already shown 
in Fig.~\ref{fig:denscomp}. For both of them we found very good agreement with the reference data and the 
crucial question now is if this translates also into the corresponding physical 
observable matching the reference data well. In Fig.~\ref{fig:condensate} we show the CanDos results (red dots)
for the condensate $\langle \, \overline{\psi}(x) \psi(x) \, \rangle_N$ as a function of the net quark 
number $N$. Indeed we find a very satisfactory agreement with the results from Fourier transformation (black
diamonds) up to $N = 7$ where the first deviations become visible. Again, for higher values of $N$ a
more precise determination of the involved densities will be necessary.

We close the discussion of our numerical test by stressing once more, that the results presented 
here can only be considered a very preliminary assessment of the new CanDos approach. The tests were done 
in two dimensions and only the free case was considered (although this already constitutes a non-trivial test of
the method). Currently we are extending the assessment of CanDos by implementing a study in 2-d QCD, but also 
started to explore lattice field theories with 4-fermi interactions. 

\section{Direct grand canonical DoS approach}

In this section we now briefly discuss our second DoS approach which is based on a suitable pseudo-fermion 
representation of the grand canonical QCD partition sum (GCDoS approach). 
We will determine the imaginary part of the 
pseudo-fermion action and set up the FFA to compute the density as a function of the imaginary part. 

\subsection{Pseudo-fermion representation and introduction of densities}

The starting point is the grand canonical partition sum of QCD. We again consider two flavors
of staggered fermions such that the grand canonical partition sum at chemical potential $\mu$ is given by   
\begin{equation}
Z_\mu \; = \; \int \! \! \mathcal{D}[U]\;
e^{\,-S_G[U]} \; \det D[U,\mu]^{\,2} \; ,
\label{Zgc}
\end{equation}
where $S_G[U]$ is again the Wilson gauge action (\ref{Sgauge}), and the staggered Dirac operator
$D[U,\mu]$ is specified in (\ref{dirac_op}). 

We first identically rewrite the fermion determinant and subsequently express the part with the 
complex action problem in terms of pseudo-fermions,  
\begin{equation}
\det D [U,\mu] \; = \; 
\det(D[U,\mu] D[U,\mu]^\dagger ) \frac{1}{\det D[U,\mu]^\dagger }
\; = \; C \, \det(D[U,\mu] D[U,\mu]^\dagger ) \int \!\! \mathcal{D}[\phi] \; e^{\, -\phi^\dagger D[U,\mu]^\dagger\phi} \; ,
\end{equation}
where $C$ is an irrelevant numerical constant and 
$\phi(x)$ are complex-valued pseudo-fermion fields that have 3 color components. 
The measure $\int \mathcal{D}[\phi]$ simply is a product measure where at every site 
of the lattice each component is integrated over the complex plane. The overall factor  
$\det(D[U,\mu] D[U,\mu]^\dagger)$ is obviously real and positive and can be treated with 
standard techniques \cite{Cheb1,Cheb2}. The exponent of the pseudo-fermion integral on the other hand has a 
non-vanishing imaginary part and thus requires a strategy for dealing with the corresponding 
complex action problem.   

To set up the direct DoS approach in the grand canonical formulation 
we divide the exponent of the pseudo-fermion path integral into real and imaginary parts,
\begin{equation}
\phi^\dagger D[U,\mu]^\dagger\phi \, = \, S_R[\phi,U,\mu] \, - \, i \, X[\phi,U,\mu] \; , \; \; 
S_R[\phi,U,\mu] \,  = \, \phi^\dagger A[U,\mu]\phi \; ,  \; \;
X[\phi,U,\mu] \,  = \, \phi^\dagger B[U,\mu] \phi \; ,
\end{equation}
where we have defined two matrices for the kernels of the real and imaginary parts of the pseudo-fermion action,
\begin{equation}
A[U,\mu] \; = \;  \frac{D[U,\mu] + D[U,\mu]^\dagger}{2} \; , \; \; 
B[U,\mu] \; = \; \frac{D[U,\mu] - D[U,\mu]^\dagger}{2i}\;.
\end{equation}
It is straightforward to evaluate $A[U,\mu]$ and $B[U,\mu]$ explicitly,
\begin{eqnarray}
A[U,\mu]_{x,y} & = & m\delta_{x,y}\mathds{1} + \frac{1}{2}\sum_{\nu=1}^d\eta_\nu(x)\sinh(\mu\delta_{\nu,d})
\bigg[U_\nu(x)\, \delta_{x+\hat{\nu},y} \, + \, U^\dagger_\nu(x-\hat\nu) \, \delta_{x-\hat{\nu},y}\bigg]\; ,
\nonumber 
\\
B[U,\mu]_{x,y} & = & -\frac{i}{2}\sum_{\nu=1}^d\eta_\nu(x)\cosh(\mu\delta_{\nu,d})
\bigg[U_\nu(x) \, \delta_{x+\hat{\nu},y} \, - \, U^\dagger_\nu(x-\hat\nu) \, \delta_{x-\hat{\nu},y}\bigg]\;.
\end{eqnarray}
The fermion determinant thus assumes the form
\begin{equation}
\det D[U,\mu] \; = \; C \, \det(D[U,\mu] D[U,\mu]^\dagger ) \int\!\! \mathcal{D}[\phi] \; e^{\,-S_R[\phi,U] \, + \, iX[\phi,U]} \; .
\end{equation}
We have already remarked that the real and positive overall factor 
$\det(D[U,\mu] D[U,\mu]^\dagger )$ can be treated with conventional simulation techniques \cite{Cheb1,Cheb2}
which we will not address in detail here (see \cite{GaMaTo} for a discussion of this term in the 
Wilson fermion formulation). Together with the Boltzmann factor for the gauge field action we combine 
this term into a new effective action Boltzmann factor defined as
\begin{equation}
e^{\,-S_{eff}[U,\mu]} \; = \; e^{\,-S_G[U]} \, \det(D[U,\mu]D[U,\mu]^\dagger) \; .
\end{equation}
The grand-canonical partition sum thus can be written as
\begin{equation}
Z_\mu \; = \; \int\!\!\mathcal{D}[U]\int\!\!\mathcal{D}[\phi] \; e^{\, -S_{eff}[U,\mu]} \; 
e^{\, - S_R[\phi,U,\mu]} \; e^{\, i \, X[\phi,U,\mu]} \; .
\end{equation}
The next step is to introduce suitable densities for the imaginary part
\begin{equation}
\rho^{^{(J)}}\!(x) \; = \; \int\!\!\mathcal{D}[U]\int\!\!\mathcal{D}[\phi] \; e^{\, -S_{eff}[U,\mu]} \; 
e^{\, - S_R[\phi,U,\mu]} \; J[\phi,U,\mu] \; \delta\big( x \, - \, X[\phi,U,\mu] \big) \; ,
\end{equation}
where we again allow for the insertion of functionals $J[\phi,U,\mu]$ in order to take into account different 
observables. As for the CanDos approach one may use charge conjugation symmetry to show that the densities
are either even or odd functions of $x$, depending on the insertion $J[\phi,U,\mu]$ (see \cite{GaMaTo}). 
Thus it is sufficient to compute the densities only for positive $x$. 

With the help of the densities vacuum expectation values of observables in the grand canonical picture at 
chemical potential $\mu$ can be written as
\begin{equation}
\langle {\cal O} \rangle_\mu \; = \; \frac{1}{Z_\mu} \int_0^\infty dx \;
\rho^{^{({\cal O})}}\!(x) \; e^{\, i x} \; , \; \;  
Z_\mu \; = \; \int_0^\infty dx \;
\rho^{^{(\mathds{1})}}\!(x) \; e^{ \, i x} \; . 
\label{obsmu}
\end{equation}

\subsection{Evaluation of the densites with FFA}

Having defined the densities and expressed grand canonical vacuum expectation values as suitable 
integrals over the densities we now can set up the FFA approach for evaluating the densities. 

First we remark 
that the densities $\rho^{^{(J)}}\!(x)$ are expected to be fast decreasing functions of $x$ and in 
\cite{GaMaTo} this has indeed been verified in test cases. Thus we may cut off the integration range 
in (\ref{obsmu}) to a finite interval $[0,x_{max}]$ and determine the density only for this range. 
As for the canonical case we divide the interval $[0,x_{max}]$ into $M$ intervals $I_n = [x_n,x_{n+1}]$,
$n = 0,1, \, ... \; M-1$, with $x_0 = 0$ and $x_M = x_{max}$.
As for the CanDos formulation the densities are parameterized by the negative exponential of a function $L(x)$  
that is continuous and piecewise linear on the intervals $I_n$. Again we assume the normalization $L(0) = 0$
and the density thus is entirely determined by the slopes $k_n$.

In the FFA approach the slope $k_n$ in each interval $I_n$ is determined from suitable restricted vacuum 
values which we here define as
\begin{equation}
\langle X\rangle_n(\lambda) \; = \; 
\frac{1}{Z_n(\lambda)}\int\mathcal{D}[U]\int\mathcal{D}[\phi] \, e^{\, -S_{eff}[U,\mu]} \; 
e^{\, - S_R[\phi,U,\mu]} \; e^{\, \lambda \, X[\phi,U,\mu] } \; J[\phi,U,\mu] \; \Theta_n\big(X[\phi,U,\mu]\big) \; ,
\label{vevrestmu}
\end{equation}
where we have defined the support function $\Theta_n(x)$
\begin{equation}
\Theta_n(x) \; = \; \left\{ \begin{array}{c}
 1 \; \mbox{for} \; x \, \in \, I_n \; , \\
 0 \; \mbox{else} \; . 	
\end{array}
\right. 
\end{equation}
As in the canonical case also the generalized expectation values (\ref{vevrestmu}) can be expressed in terms
of the parameterized density and computed in closed form, along the lines discussed above. 
After normalizing them to the form (\ref{vtilde})
the generalized expectation values are again described by the functions  
$h\big(\Delta_n[\lambda-k_n]\big)$ such that the slopes $k_n$ can be determined from 1-parameter fits. 
Subsequently the densities are constructed from the slopes using (\ref{piecewise}) and (\ref{rho_interval}),
with $\theta$ replaced by $x$. Finally we can compute observables from the densities using (\ref{obsmu}). 

The direct, grand canonical density of states approach discussed in this section for staggered fermions
has been discussed for Wilson fermions in \cite{GaMaTo}. There also first exploratory numerical results
were presented and for free fermions it was shown that the density obtained with the FFA approach matches 
exact reference data from Fourier transformation very well.  


\section{Summary and outlook}

In this paper we have extended our previous work \cite{GaMaTo}, where we presented two new DoS techniques for 
finite density lattice QCD with Wilson fermions, to the formulation of QCD with staggered fermions. The first
formulation is based on the canonical formulation where the canonical partion sum and vacuum expectation values
of observables at fixed net quark number are obtained as Fourier moments with respect to imaginary chemical 
potential. The functional fit approach (FFA) can then be used to compute the density with sufficient accuracy
for reliably determining observables for reasonable net quark numbers. We present exploratory tests of the
canonical DoS approach for the case of free fermions in 2-d and find that observables such as the chiral condensate
at finite net quark numbers reliably match reference data obtained from a direct calculation with Fourier 
transformation that is possible in the free case.

Our second approach is set up directly in the grand canonical ensemble. The QCD partition sum is rewritten in terms
of a suitable pseudo-fermion representation and the imaginary part of the pseudo-fermion action is identified.
Using FFA the density is then computed as a function of the imaginary part and grand canonical vacuum expectation 
values are again obtained as the corresponding oscillating integrals.

Two comments are in oder here: Although the first tests are encouraging, the numerical results presented 
here clearly constitute only a very preliminary and exploratory assessment of the new techniques. We are currently
extending these tests towards QCD in two dimensions as the next test case before approaching the full 4-d theory.
We furthermore stress that the techniques we have presented here are not restricted to QCD or other gauge theories
with fermions. Also theories with 4-fermi interactions can be accessed after the introduction of suitable
Hubbard Stratonovich fields and also for this direction of possible further development we have started 
exploratory calculations.        

\vskip8mm
\noindent
{\bf Acknowledgments:} We thank Mario Giuliani, Peter Kratochwill, Kurt Langfeld and  Biagio Lucini
for interesting discussions. This work is supported by the Austrian Science Fund FWF, grant I 2886-N27 and 
partly also by the FWF DK 1203 ''Hadrons in Vacuum Nuclei and stars''.



\end{document}